\documentclass[12pt]{emulateapj}
\usepackage{epsfig} \usepackage{amsmath} \usepackage{amssymb}
\usepackage{natbib} \usepackage{graphicx}

 \def\kpc{{\rm\thinspace kpc\ }}

\shorttitle{Size and dispersion evolution of early-type galaxies}
\shortauthors{Oser et al.}

\begin{document}

\title{The cosmological size and velocity dispersion evolution of
  massive early-type galaxies}

\author{Ludwig Oser$^1$$^2$, Thorsten Naab$^1$, Jeremiah
  P. Ostriker$^3$, Peter H. Johansson$^4$$^5$} \affil{$^1$
  Max-Planck-Institut f\"ur Astrophysik, Karl-Schwarzschild-Strasse 1,
  85741, Garching, Germany \\ $^2$ Universit\"ats-Sternwarte
  M\"unchen, Scheinerstr.\ 1, D-81679 M\"unchen, Germany \\ $^3$
  Department of Astrophysical Sciences, Princeton University,
  Princeton, NJ 08544, USA\\ $^4$ Department of Physics, University of
  Helsinki, Gustaf H\"allstr\"omin katu 2a, FI-00014 Helsinki, Finland
  \\ $^5$ Finnish Centre for Astronomy with ESO, University of Turku,
  V\"ais\"al\"antie 20, FI-21500 Piikki\"o, Finland
  \\ \texttt{oser@mpa-garching.mpg.de} \\}

\begin{abstract}
We analyze 40 cosmological re-simulations of individual massive
galaxies with present-day stellar masses of $M_{*} > 6.3 \times
10^{10} M_{\odot}$ in order to investigate the physical origin of the
observed strong increase in galaxy sizes and the decrease of the
stellar velocity dispersions since redshift $z \approx 2$.  At present
25 out of 40 galaxies are quiescent with structural parameters (sizes
and velocity dispersions) in agreement with local early type galaxies.
At z=2 all simulated galaxies with $M_* \gtrsim 10^{11}M_{\odot}$ (11
out of 40) at z=2 are compact with projected half-mass radii of
$\approx$ 0.77 ($\pm$0.24) kpc and line-of-sight velocity dispersions
within the projected half-mass radius of $\approx$ 262 ($\pm$28)
kms$^{-1}$ (3 out of 11 are already quiescent).  Similar to observed
compact early-type galaxies at high redshift the simulated galaxies
are clearly offset from the local mass-size and mass-velocity
dispersion relations. Towards redshift zero the sizes increase by a
factor of $\sim 5-6$, following $R_{1/2} \propto (1+z)^{\alpha}$ with
$\alpha = -1.44$ for quiescent galaxies ($\alpha = -1.12$ for all
galaxies). The velocity dispersions drop by about one-third since $z
\approx 2$ , following $\sigma_{1/2} \propto (1+z)^{\beta}$ with
$\beta = 0.44$ for the quiescent galaxies ($\beta = 0.37$ for all
galaxies). The simulated size and dispersion evolution is in good
agreement with observations and results from the subsequent accretion
and merging of stellar systems at $z\lesssim 2$ which is a natural
consequence of the hierarchical structure formation. A significant
number of the simulated massive galaxies (7 out of 40) experience no
merger more massive than 1:4 (usually considered as major mergers). On
average, the dominant accretion mode is stellar minor mergers with a
mass-weighted mass-ratio of 1:5. We therefore conclude that the
evolution of massive early-type galaxies since $z \approx 2$ and their
present-day properties are predominantly determined by frequent
'minor' mergers of moderate mass-ratios and not by major mergers
alone.
\end{abstract}


\section{Introduction}
There is growing observational evidence for the existence of a
population of massive galaxies ($\approx 10^{11} M_{\odot}$) with
small sizes ($\approx 1 \; \rm kpc$) and low star formation rates at
redshift $z \ge 2$.  These galaxies are smaller by a factor of three
to five compared to present-day ellipticals at similar masses
(e.g. \citealp{2009MNRAS.394.1978H}) and their effective stellar
densities are at least one order of magnitude higher
\citep{2005ApJ...626..680D,2006ApJ...650...18T,2007MNRAS.374..614L,
  2007ApJ...671..285T,2007MNRAS.382..109T,2007ApJ...656...66Z,
  2008ApJ...687L..61B,2008ApJ...677L...5V,2008A&A...482...21C,
  2008ApJ...688..770F, 2009MNRAS.392..718S, 2009ApJ...697.1290B,
  2009ApJ...695..101D, 2011arXiv1108.0656D}. Deep observations down to
low surface brightness limits (H $\approx$ 28 mag arcsec$^{-2}$) show
no evidence for faint, previously missed, stellar envelopes
\citep{2010MNRAS.405.2253C,2010ApJ...714L.244S} and measurements of
higher velocity dispersions seem to independently confirm previous
high mass estimates \citep{2005ApJ...631..145V,2009ApJ...698.1232V,
  2009ApJ...696L..43C, 2009Natur.460..717V, 2009ApJ...704L..34C,
  2011ApJ...738L..22M}.

Quiescent (red \& dead) galaxies make up about half of the general
high redshift (z $\approx$ 2) population of massive galaxies and most
of them (90\%) are found to be compact
\citep{2006ApJ...649L..71K,2006ApJ...638L..59V,
  2008ApJ...677L...5V,2009ApJ...691.1879W}. In the local Universe,
galaxies of similar mass and size are extremely rare
\citep{2009ApJ...692L.118T} or do not exist at all
\citep{2010ApJ...720..723T}. This indicates that present-day
early-type galaxies were not fully assembled at $z\approx 2$ and
underwent significant structural evolution until the present
day. Observations of the growth of massive galaxies since $z \approx
2$ selected at constant number density \citep{2010ApJ...709.1018V}
indicate that they grow inside-out. A quiescent - without significant
in-situ formation of new stars - build-up of extended stellar
envelopes can originate from minor mergers and was predicted from
cosmological simulations \citep{2007ApJ...658..710N,
  2009ApJ...699L.178N, 2010ApJ...725.2312O} and recently, for the
first time such minor mergers at high redshift might have been
directly observed \citep{2010ApJ...718L..73V,2010MNRAS.405.2253C}.

A simple picture of high redshift monolithic formation or, similarly,
a binary merger of massive very gas-rich disks at z $\gtrsim$ 2 -
which has been suggested as a reasonable formation mechanism for
compact high-redshift galaxies
\citep{2010ApJ...722.1666W,2010MNRAS.406..230R,2011ApJ...730....4B} -
followed by passive evolution can be ruled out
\citep{2008ApJ...682..896K,2008ApJ...677L...5V,2009ApJ...692L.118T}
unless the increase in size of ellipticals can be explained by secular
processes such as adiabatic expansion driven by stellar mass loss
and/or strong feedback
\citep{2008ApJ...689L.101F,2009ApJ...695..101D,2010ApJ...718.1460F}.
This process seems to be disfavored by observations
(e.g. \citealp{2010MNRAS.401.1099H, 2011MNRAS.415.3903T}) and the
absence of a significant young stellar population would indicate that
such hypothesized secular processes would need to occur without
significant star formation.

Based on high-resolution cosmological simulations of individual
galaxies, \citet{2007ApJ...658..710N,2009ApJ...699L.178N} and
\citet{2010ApJ...725.2312O} provide an explanation for the size growth
and the decrease in velocity dispersion, which is consistent with the
cosmological hierarchical buildup of galaxies.  The compact cores of
massive galaxies form during an early rapid phase of dissipational
in-situ star formation at $6 \gtrsim z \gtrsim 2$ fed by cold flows
\citep{2005MNRAS.363....2K,2009Natur.457..451D,2010ApJ...725.2312O}
and/or gas rich mergers {leading to large stellar surface
  densities \citep{2011arXiv1107.2591W}}.  At the end of this phase
the observed as well as simulated galaxies are more flattened and
disk-like than their low redshift counterparts
\citep{2008ApJ...677L...5V,2009ApJ...699L.178N,2011ApJ...730...38V}.
They are already massive ($\approx 10^{11}M_{\odot}$) but have small
sizes of $\approx$ 1 kpc and velocity dispersions of $\approx$ 240
kms$^{-1}$ (see also \citealp{2009ApJ...692L...1J}), in general
agreement with observations.  The subsequent evolution is dominated by
the addition of stars that have formed ex-situ, i.e. outside the
galaxy itself \citep{2010ApJ...725.2312O}. {These accreted
  stars typically settle at larger radii \citep[see
    also][]{2011arXiv1108.3834C}}.  The early domination of in-situ
star formation and the subsequent growth {by stellar mergers}
is in agreement with predictions from semi-analytical models
\citep{1996MNRAS.281..487K,2006ApJ...648L..21K,
  2006MNRAS.366..499D,2007MNRAS.375....2D,2008MNRAS.384....2G,
  2010MNRAS.405..948S, 2010MNRAS.403..117S} {and the assembly
  scenario discussed in \citet{2009ApJS..182..216K}}.

{In the absence of gas,} stellar (i.e. collisionless) accretion
and 'dry' merging in general is an energy conserving process {in the
  sense that none of the gravitational and binding energy in the
  accreted systems can be radiated away during the merging event.}
Therefore, while the galaxies grow in mass, they must significantly
increase their sizes and, eventually, decrease their velocity
dispersions during this phase, in particular if the stars are accreted
in minor mergers. {In massive galaxies that are embedded in a
  hot gaseous halos some fraction of the gravitational energy can be
  radiated away \citep{2009ApJ...697L..38J}}.
\citet{2000MNRAS.319..168C}, \citet{2009ApJ...699L.178N} and
\citet{2009ApJ...697.1290B} presented the simple virial arguments for
why minor mergers lead to a stronger size increase and a decrease in
velocity dispersion than the more commonly studied major mergers
\citep{2006ApJ...636L..81N}.

In \citet{2010ApJ...725.2312O} we investigated this two phase scenario
in more detail with a larger sample of re-simulations and found a
connection between galaxy mass, size, and the assembly history (see
also
\citealp{2006ApJ...648L..21K,2007MNRAS.375....2D,2008MNRAS.384....2G,
  2009ApJ...691.1424H,2010ApJ...709..218F}). More massive present-day
systems contain a larger fraction of accreted stars (up to 80 per
cent) which, over time, build an outer envelope and increase the size
of the systems \citep{2009ApJ...699L.178N,2009MNRAS.398..898H,
  2010ApJ...725.2312O, 2010ApJ...709..218F}. This scenario receives
support from recent observational findings that massive galaxies have
increased their mass at radii $r > 5 \, \rm{kpc}$ by a factor of
$\approx$ 4 since z=2 with the mass at smaller radii being essentially
unchanged \citep{2010ApJ...709.1018V}.  {It is also the favored
  model to explain observed kinematics \citep{2011ApJ...736L..26A} and
  metallicity gradients \citep{2011MNRAS.413.2943F} of globular
  cluster populations in nearby elliptical galaxies.}

In this paper we analyze a subset of massive galaxies from the
\citet{2010ApJ...725.2312O} simulations with a particular focus on the
evolution of sizes and velocity dispersions since $z \approx 2$. The
paper is organized as follows: In section \ref{SIMULATIONS} we briefly
review the simulations. The results on the evolution of size and
velocity dispersion are presented in sections \ref{size} and
\ref{dispersion}. The stellar merger histories of the resimulated
galaxies are reviewed in section \ref{merger}.  We conclude and
discuss our results in section \ref{discussion}.

\section{High resolution simulations of individual galaxy halos} 
\label{SIMULATIONS}

The results presented in this paper are drawn from 40 'zoom-in'
hydrodynamic simulations of individual halos which are presented in
detail in \citet{2010ApJ...725.2312O}. The halos are picked from a
dark matter only simulation using a flat cosmology with parameters
obtained from WMAP3 \citep{2007ApJS..170..377S}: $\mathrm{h}=0.72, \;
\Omega_{\mathrm{b}}=0.044, \; \Omega_{\mathrm{dm}}=0.216, \;
\Omega_{\Lambda}=0.74, \; \sigma_8=0.77 $ and an initial slope of the
power spectrum of $\mathrm{n_s}=0.95$. From redshift zero we trace
back in time all particles close to the halos of interest at any given
snapshot. Those particles are then replaced with high-resolution gas
and dark matter particles.  The original dark matter particles are
merged (depending on their distance to the re-simulated halo) to
reduce the particle count and the simulation time.  The new high mass
resolution initial conditions are evolved from redshift z=43 to the
present day using a modified version of the parallel TreeSPH code
GADGET-2 \citep{2005MNRAS.364.1105S} including star formation,
supernovae feedback \citep{2003MNRAS.339..289S} and cooling for a
primordial composition of hydrogen and helium.  Additionally, the
simulations include a redshift-dependent UV background radiation field
with a modified \citealp{1996ApJ...461...20H} spectrum.

The simulated halo masses cover the range $7 \times 10^{11}
M_{\odot}h^{-1}$ $ \lesssim M_{\mathrm{vir}} \lesssim 2.7 \times
10^{13} M_{\odot}h^{-1}$ and the central galaxy masses are between
$4.5 \times 10^{10} M_{\odot}h^{-1} \lesssim M_* \lesssim 3.6 \times
10^{11} M_{\odot}h^{-1}$ at $z=0$. The masses for the gas and star
particles are $m_{*,gas}=4.2 \times 10^{6}M_{\odot}h^{-1}$ (we spawn
one star particle per gas particle), whereas the dark matter particles
have a mass of $m_{\mathrm{dm}} = 2.5 \times 10^{7}M_{\odot}h^{-1}$.
The comoving gravitational softening lengths used are
$\epsilon_{\mathrm{gas,star}} = 400 \rm pc \, h^{-1}$ for the gas and
star particles and $\epsilon_{\mathrm{halo}} = 890 \rm pc \, h^{-1}$
for the dark matter. At $z \approx 2$ the corresponding physical
softening length for stars is $\epsilon_{\mathrm{gas,star}} = 133 \rm
pc \, h^{-1}$. The integration accuracy parameter was set to 0.005 to
guarantee accurate time integration \citep{2005MNRAS.364.1105S}. In
the following we present the results for 40 galaxies with masses
larger than $M_* \approx 6.3 \times 10^{10} M_{\odot}$ for direct
comparison with observations. These galaxies are well resolved with
$\approx 1.5 \times 10^5 - 3 \times 10^6$ particles within the virial
radius ($R_{\mathrm{vir}} \equiv R_{200}$, the radius where the
spherical overdensity drops below 200 times the critical density of
the universe at a given redshift).  Using the above simulation
parameters for zoom simulations have been shown to result in galaxies
with reasonable present-day properties
\citep{2007ApJ...658..710N,2009ApJ...697L..38J,2009ApJ...699L.178N,2010ApJ...725.2312O}. However,
the fraction of available baryons converted into stars, $f_*$, for
galaxies in this mass range is typically 2 times higher than estimates
from models that are constructed by matching observed luminosity
functions to simulated halo mass functions \citep[and references
  therein]{2010MNRAS.404.1111G,2010ApJ...710..903M,2010ApJ...717..379B}.

\begin{figure*}
\centering
\includegraphics[width=18cm]{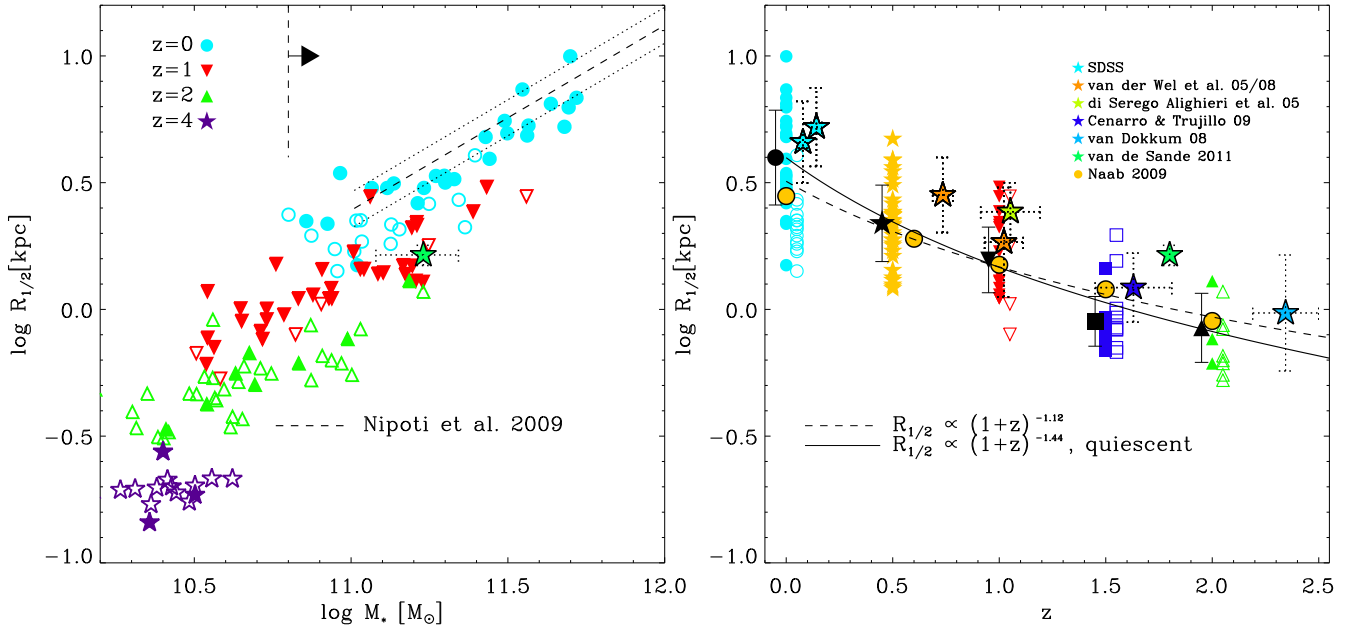}
\caption{Left: Projected stellar half-mass radii of the simulated
  galaxies versus stellar masses for redshifts z=0 (blue circles), z=1
  (red triangles), z=2 (green triangles) and z=4 (purple
  stars). Quiescent galaxies with low specific star formation rates
  ($sSFR \leq 0.3/t_{\mathrm{H}}$) have solid symbols, other galaxies
  have open symbols. Since $z \approx 2$ all galaxies evolve rapidly
  in size. The dashed line indicates the observed size-mass relation
  for early-type galaxies of \citep{2009ApJ...706L..86N} {with the
    one-sigma scatter indicated by the dotted lines}. The z=0 mass cut
  of $M_* > 6.3 \times 10^{10}M_{\odot}$ for the galaxy sample is
  indicated by the vertical dashed line. Right: Projected stellar
  half-mass radii of galaxies with stellar masses $M_* > 6.3 \times
  10^{10}M_{\odot}$ (see arrow on the left plot) as a function of
  redshift. The black symbols indicate the mean sizes at a given
  redshift with the error bars showing the standard deviation. The
  star forming galaxies (open symbols) and mean values are offset by
  0.1 in redshift for clarity. The black lines show the result of a
  power law fit for all (dotted line) and quiescent (solid line)
  systems, respectively in good agreement with observed relations.
  Observational estimates from different authors are given by the
  solid star symbols where the dotted error bars show the observed
  scatter (see \citealp{2008ApJ...677L...5V,2009ApJ...696L..43C})
  {with the exeption of the observation by
    \citet{2011ApJ...736L...9V}. Since this is a single object, here
    the error bars indicate the uncertainty of the measurement.}. By
  z=3 all progenitor galaxies drop below our mass limit.  }
\label{rad-gm-z0-shaded}
\end{figure*}

\begin{figure}
\centering
\includegraphics[width=8.5cm]{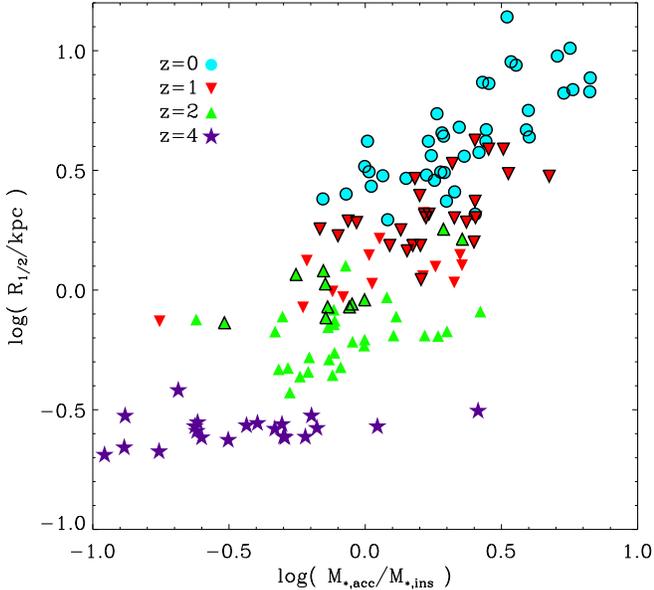}
\caption{Projected stellar half-mass radius of the simulated galaxies
  at different redshifts (see Fig. \ref{rad-gm-z0-shaded}) versus the
  fraction of stellar mass accreted (in major mergers, minor mergers
  and accretion events), $M_{*,\mathrm{acc}}$, to the stellar mass
  formed in-situ, $M_{*,\mathrm{ins}}$, in the galaxies. The black
  bordered symbols indicate systems more massive than $M_* > 6.3
  \times 10^{10}M_{\odot}$. At $z \gtrsim 2$ galaxies with a higher
  fraction of accreted stars have larger sizes indicating that
  accretion of stellar systems drives the size evolution of massive
  galaxies.}
\label{ia-rad-z}
\end{figure}

\begin{figure}
\centering
\includegraphics[width=8.5cm]{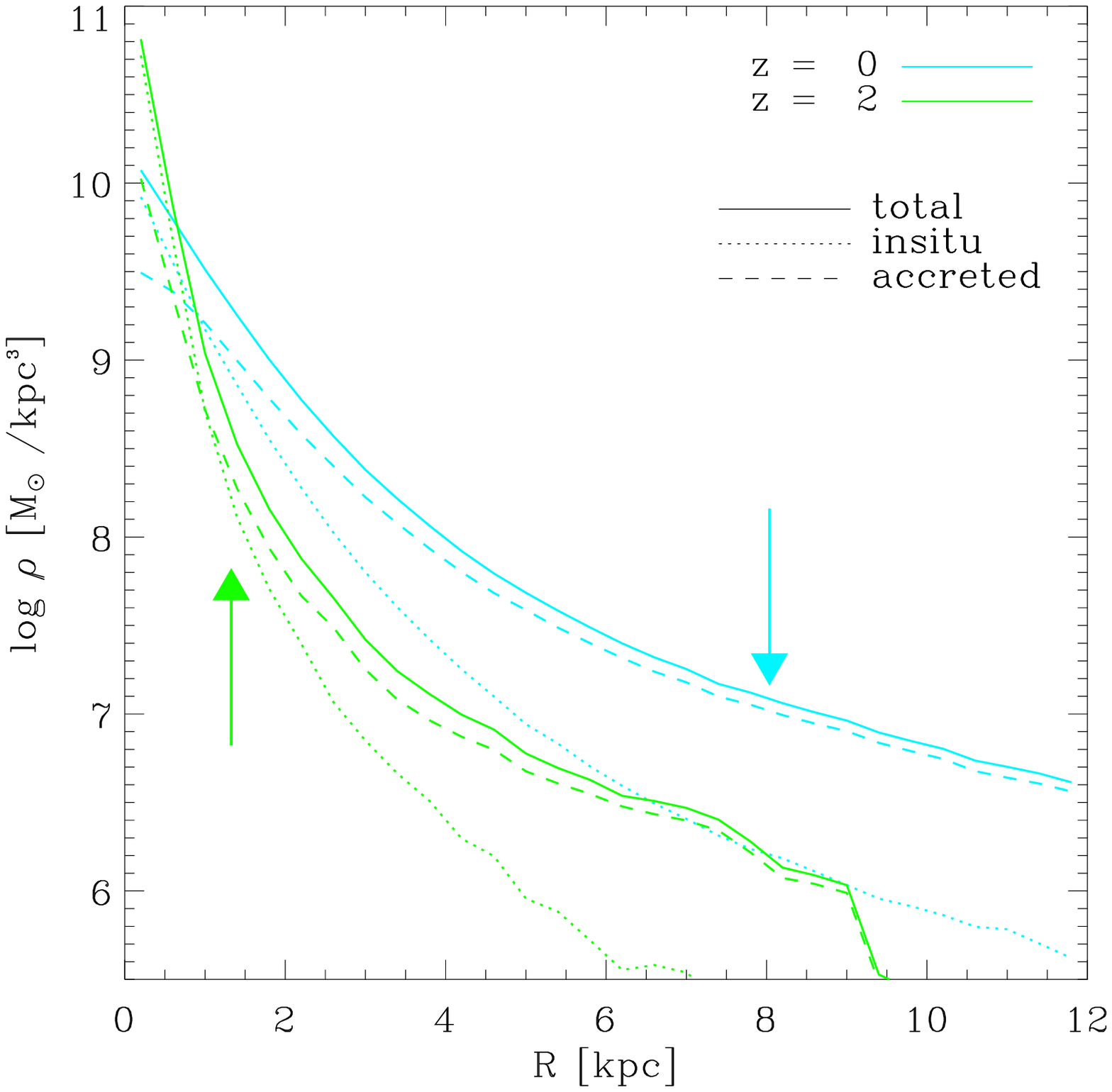}
\caption{Stellar density as a function of
  radial distance to the galactic center at redshift 0 (blue) and 2
  (green) averaged over the most massive systems ($M_{*} > 2\times
  10^{11}M_{\odot}$ at $z=0$). The arrows indicate the average half-mass radii at a given
  redshift. At $z=2$ we find that the stellar mass profile inside
  the half-mass radius is still dominated by stars that have formed
  in-situ (dotted lines). At $z=0$ the half-mass radii of our galaxies have
  significantly extended due to accreted stars (dashed lines) which dominate
  the total density (solid lines) at larger radii.}
\label{rad-density}
\end{figure}

\section{Redshift evolution of sizes} 
\label{size}



In the left panel of Fig. \ref{rad-gm-z0-shaded} we show the projected
half-mass radii of the simulated galaxies as a function of their
stellar mass at z=0 (blue circles), as well as the sizes and masses of
their most massive progenitors at z=1 (red triangles), z=2 (green
triangles), and z=4 (purple stars).  {We determine the center
  of the galaxies using the shrinking sphere technique described in
  \citet{2003MNRAS.338...14P}, starting at redshift 0 with a sphere
  that contains all the stellar particles. In all the previous
  snapshots the center of the most massive progenitor is found with
  the same technique starting with a sphere of a radius of at least
  $25 \kpc$ that encompasses the 50 innermost particles identified in
  the last processed snapshot.}  The sizes indicated here are the mean
values of the half-mass radii of all stars within $0.1 \times
R_{\mathrm{vir}} \equiv R_{\mathrm{Gal}}$ (considered the central
galaxy) projected along the three principal axes of the main stellar
body.  {We always show the median of the sizes of the galaxies
  that we compute from the snapshot at the target redshift as well as
  the two snapshots before and after this one to avoid outliers caused
  interacting systems.}  We separate the sample into quiescent
galaxies (solid symbols) with specific star formation rates
$\mathrm{sSFR} < 0.3/t_{\mathrm{H}}$ \citep{2008ApJ...688..770F},
where $t_{\mathrm{H}}$ is the age of the Universe at each
redshift. Star forming galaxies are indicated by open symbols and have
$\mathrm{sSFR} > 0.3/t_{\mathrm{H}}$. The dashed black line shows the
z=0 linear fit to the SLACS sample of local early-type galaxies
\citep{2009ApJ...706L..86N} {with the observed scatter given by
  the dotted lines}, which is in good agreement with the simulated
galaxy sizes. Other published local mass-size relations have slightly
different slopes and offsets (see
e.g. \citealp{2009MNRAS.394.1978H,2009MNRAS.398.1129G,2010ApJ...713..738W})
which does, however, not affect our general conclusions. At z=4 all
progenitor galaxies are very compact with similar sizes ($\approx
300\,\rm pc$) independent of their mass. During this phase the
formation of the proto-galaxies is dominated by gas dissipation and
in-situ star formation
(\citealp{2009ApJ...699L.178N,2009ApJ...692L...1J,2010ApJ...725.2312O}). By
z=2 a clearly visible mass-size relation has already developed. At
this epoch the most massive galaxies of our sample have experienced
the most rapid size growth with half-mass radii up to $\approx 1.3 \,
\rm kpc$ for galaxies with $10^{11} M_{\odot}$ in stellar mass, in
good agreement even with the most recent observations
(e.g. \citealp{2010arXiv1007.1460R, 2011ApJ...730...38V,
  2011arXiv1106.4308C}).  Towards z=0 the simulated galaxies continue
to grow in size as well as mass. The descendants of galaxies that are
already massive ($>6.3 \times 10^{10}M_{\odot}$) at z=2 (green symbols
to the right of the vertical dashed line) increase their mass by a
factor of 3.5 ($77 \pm 10 \%$ of the accumulated mass is due to
stellar accretion) and their projected half mass radii grow by a
factor of 6.5. On average, all simulated galaxies more massive than
$6.3 \times 10^{10}M_{\odot}$ at any given redshift grow by a factor
of 2.1 in mass (see e.g. \citet{2007ApJ...665..265F}) and a factor of
4.6 in radius since z=2. This already indicates that the size growth
cannot be the result of equal-mass dry mergers, since the ensuing size
growth should be, at most, directly proportional to the mass increase
\citep{2003MNRAS.342..501N,2009ApJ...706L..86N,2009ApJ...699L.178N}.
Overall, the size growth is differential, i.e. the most massive
galaxies show the strongest size increase and the tilt of the
mass-size relation increases towards lower redshifts. Fitting all
galaxies with $R_{1/2} \propto M_{*}^{\delta}$ we find $\delta \approx
0.46 \pm 0.056$ at z=2 and $\delta \approx 0.69 \pm 0.064$ at z=0. For
quiescent galaxies (with worse statistics) we find no trend for
differential size growth with $\delta \approx 0.67 \pm 0.069$ at z=2
and $\delta \approx 0.65 \pm 0.090$ at z=0. This is in qualitative
agreement with recent observations by \citet{2010ApJ...713..738W} who
do not find observational indications for a differential size growth
of quiescent galaxies.


Observed sizes of massive galaxies are found to evolve as
$(1+z)^\alpha$. Depending on the selection criteria (specific star
formation rate, concentration etc.) {and observed redshift
  range} the observed values of $\alpha$ for massive ellipticals range
{from $\alpha$ = -0.75 \citep{2010ApJ...717L.103N} to $\alpha$
  = -1.62 \citep{2011arXiv1108.0656D}}. \citet{2008ApJ...688..770F}
find $\alpha \approx -0.8$ for all galaxies in this mass range ($M_{*}
\gtrsim 10^{11} M_{\odot}$) whereas for quiescent massive galaxies the
observed size evolution is faster with $-1.09 < \alpha < -1.22$.
\citet{2011arXiv1106.4308C} obtain values from $\alpha$ = -0.87 to
$\alpha$ = -1.42 depending on stellar
mass. {\citet{2008ApJ...688...48V} find a value of $\alpha
  \approx$ -0.98, when they include results from previous surveys this
  changes to $\alpha \approx$ -1.20. A similar trend ($\alpha \approx
  -1.11$) is found for UV-bright galaxies
  \citep{2011ApJ...727....5M}.}

In the right panel of Fig. \ref{rad-gm-z0-shaded} we show the size
evolution of galaxies more massive than $6.3 \times 10^{10} M_{\odot}$
(see \citealp{2008ApJ...688..770F}) since z=2. At z=4 all progenitor
galaxies drop below the threshold mass, but are still resolved by
$\approx 10^{4}$ particles. {We added the results from
  \citet{2009ApJ...699L.178N}, which were obtained with the same
  simulation code but with a softening length fixed in physical
  units. The size evolution in this case is very similar to the
  simulations that uses a fixed comoving softening length.}
{We also included various observational results which find
  slightly larger sizes at a given redshift but with a very similar
  evolution in time}.  On average there is a strong evolution in
galaxy sizes: for the $R_{1/2} \propto (1 + z)^{\alpha}$ power law fit
to all (dashed line) and only the quiescent (solid line) simulated
galaxies we find a value of $\alpha = -1.12 \pm 0.13$ and $\alpha =
-1.44 \pm 0.16$, respectively. This is in good agreement with observed
values - {which are possibly a bit offset to higher values} -
and despite our statistical limitations we consider this trend robust.
{Possible simple explanations for an offset in size within a
semi-analytical framework are discussed in
\citet{2011arXiv1105.6043S}.}

To demonstrate the physical origin for the size growth in the
simulated galaxies we show in Fig. \ref{ia-rad-z} the projected
half-mass radii presented in Fig. \ref{rad-gm-z0-shaded} at different
redshifts as a function of the ratio of stars accreted onto the galaxy
$M_{*,\mathrm{acc}}$ to the stars formed in-situ in the galaxy,
$M_{*,\mathrm{ins}}$ at the same redshifts. We consider a star
particle in the simulation as formed in-situ in the galaxy if it is
created inside $R_{\mathrm{gal}} (\equiv 0.1 \times
R_{\mathrm{vir}})$.  Black bordered symbols indicate galaxies with
stellar masses larger than $6.3 \times 10^{10} M_{\odot}$ whose size
evolution is plotted in the right panel of
Fig. \ref{rad-gm-z0-shaded}.
There is a clear correlation between the relative amount of accreted
stars and in-situ stars not only at redshift zero
\citep{2010ApJ...725.2312O} but also at high redshifts ($z \approx
2$). This indicates that stellar accretion drives the size evolution
of the systems as soon as the accreted stars start to dominate the
total mass ($M_{*,\rm{acc}}/M_{*,\rm{ins}} > 1$) at z $\approx$ 2 as
also predicted from semi-analytical modeling
\citep{2006MNRAS.370..902K}.  At earlier times the stellar mass growth
is dominated by in-situ star formation \citep{2010ApJ...725.2312O},
i.e. the stars form out of cold gas that was able to radiate away a
large fraction of its gravitational energy and thus leading to compact
systems.  The binding energy of the accreted stars, however, is
retained and will increase the total energy content of the accreting
galaxy, both by shock-heating the gas {- which then can cool
  radiatively -} as well as expanding the existing dark matter and
stellar components \citep{2009ApJ...697L..38J}.  This in general leads
to more extended systems.

In Fig. \ref{rad-density} we compare the density profiles of a
subsample of massive galaxies ($M_{*}>2 \times 10^{11}M_{\odot}$) at
redshift 2 and the present day. In agreement with
\citet{2009ApJ...699L.178N} we find that within the half-mass radius
the high redshift systems are dominated by stars that formed in-situ
while the contribution of accreted stars to the inner mass profile is
small. At the present day the stellar mass inside the effective radius
is dominated by accreted stars added at radii larger than $> 1
\kpc$. This accretion is responsible for the strong size increase
\citep{2010ApJ...725.2312O} and is in agreement with the results from
stacked imaging for massive galaxies at a constant number density that
also show an increase in surface densities predominantly in the outer
regions \citep{2010ApJ...709.1018V}.

\begin{figure*}
\centering
\includegraphics[width=18cm]{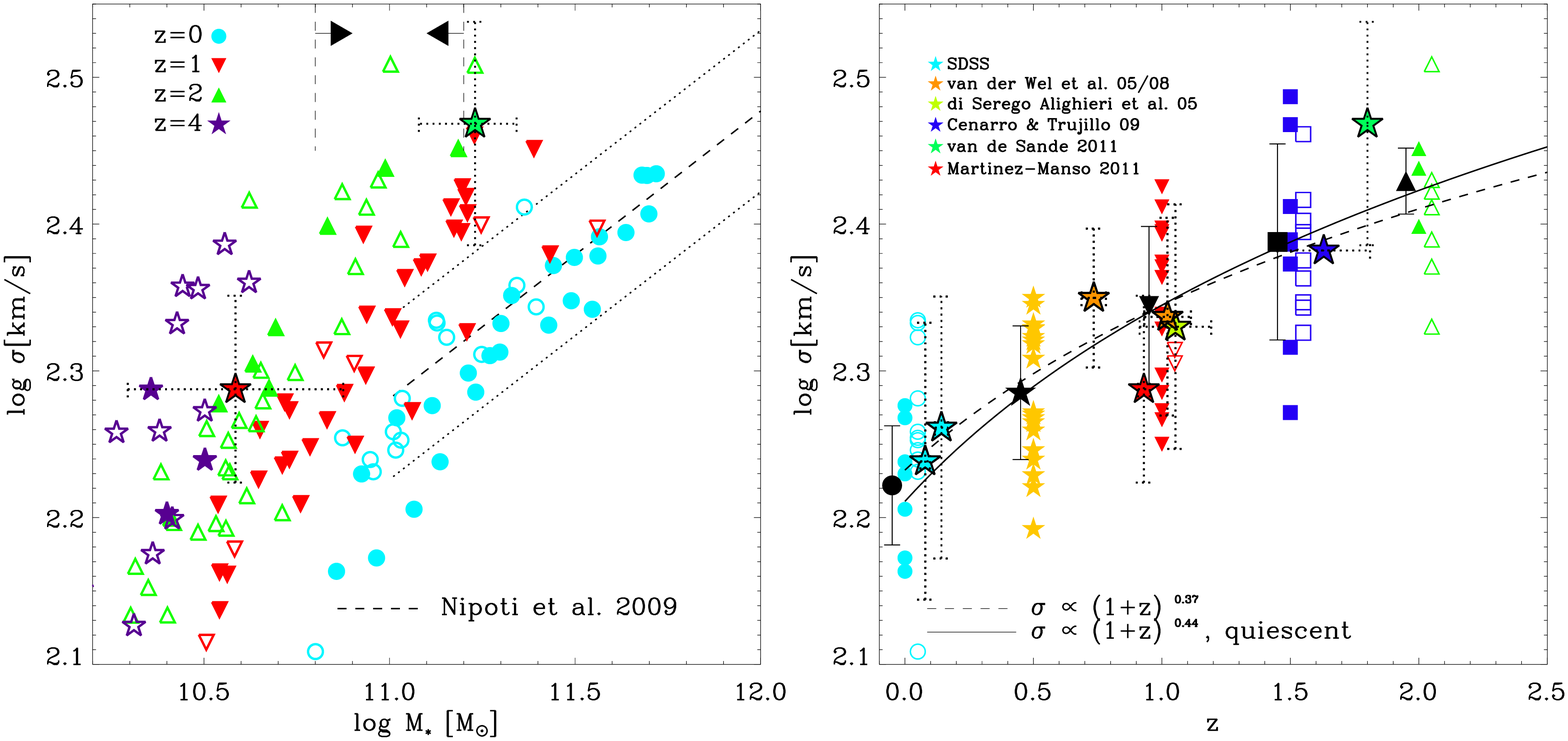}
\caption{Central (within 0.5 $R_{1/2}$) projected velocity dispersion
  as a function of stellar mass at z=0 (blue circles), z=1 (red
  triangles), z=2 (green triangles) and z=4 (purple stars). The
  relation for local galaxies from \citet{2009ApJ...706L..86N} are
  shown by the dashed line {with the dotted lines indicating the
    scatter of the observed galaxies}. At a given mass the velocity
  dispersion decreases significantly from z=4 to z=0. The mass limits
  used for the right plot are indicated by the vertical dashed lines.
  Right: Central projected velocity dispersion of the simulated
  galaxies with masses in the range of $ 6.3 \times 10^{10} M_{\odot}
  < M_{\star} < 1.6 \times 10^{11} M_{\odot} $ at any given redshift
  as a function of redshift. Solid symbols represent star forming
  galaxies and empty symbols show quiescent systems (offset by 0.1 in
  redshift for clarity). Observational estimates from different
  authors are given by the solid star symbols (see
  \citealp{2009ApJ...696L..43C,2011ApJ...736L...9V,2011ApJ...738L..22M})
          {with the observed scatter given by the dotted error bars,
            where available}.  The black lines show the result of a
          power law fit for all (dashed line) and the quiescent (solid
          line) galaxies, respectively. The simulations indicate a
          mild dispersion evolution from $\approx$ 262 kms$^{-1}$ at
          z=2 to $\approx$ 177 kms$^{-1}$ at z=0, in agreement with
          observations.}
\label{disp_mass_z}
\end{figure*}

\section{Redshift evolution of velocity dispersions}
\label{dispersion}

In the left panel of Fig. \ref{disp_mass_z} we show the central
stellar line-of-sight velocity dispersions for the simulated galaxies
as a function of the stellar mass at redshift z=4 (purple stars), z=2
(green triangles), z=1 (red triangles) and z=0 (blue circles). The
line-of-sight velocity dispersions have been calculated within $0.5
\times R_{1/2}$ along the three principal axes and then averaged. The
mass-dispersion relation from \citet{2009ApJ...706L..86N} for galaxies
more massive than $M_* \approx 10^{11} M_{\odot}$ is indicated by the
dashed line. It is evident that at a given mass range the velocity
dispersions of the galaxies systematically {increase} with
redshift. This is illustrated in the right panel of
Fig. \ref{disp_mass_z} where we show the evolution of the projected
velocity dispersion for galaxies with masses in the range of $6.3
\times 10^{11}M_{\odot} < M_{\star} < 1.6 \times 10^{12}M_{\odot}$
(indicated by the vertical lines in the left panel) as a function of
redshift. In this mass range the velocity dispersions drop from $262
\pm 28 \rm kms^{-1}$ at z=2 to $177 \pm 22 \rm kms^{-1}$ at z=0, a
decrease of roughly a factor of $1.5$. The evolution is statistically
significant but weak (see also \citealp{2009ApJ...691.1424H}). The
black lines show a fit for the average velocity dispersions for all
(dashed line) and the quiescent (solid line) galaxies only. As for the
sizes, we fit the redshift evolution of the velocity dispersions like
$\sigma_{1/2} \propto (1+z)^{\beta}$. Again we find a slightly
stronger evolution for the quiescent systems ($\beta = 0.44$) than for
all galaxies in our samples ($\beta =
0.37$). {\cite{2011ApJ...736L...9V} obtain a similar value of
  $\beta = 0.51 \pm 0.07$.} Depending on selection criteria
\citet{2010A&A...524A...6S} find values for $\beta$ ranging from $0.59
\pm 0.10$ to $0.19 \pm 0.10$. Following \citet{2009ApJ...696L..43C} we
compare to observations of local ellipticals and measurements at
higher redshift
\citep{2005A&A...442..125D,2005ApJ...631..145V,2008ApJ...688...48V,2009ApJ...696L..43C}.
      {In general we find a good agreement with the observations.}

\begin{figure}
\centering
\includegraphics[width=8.5cm]{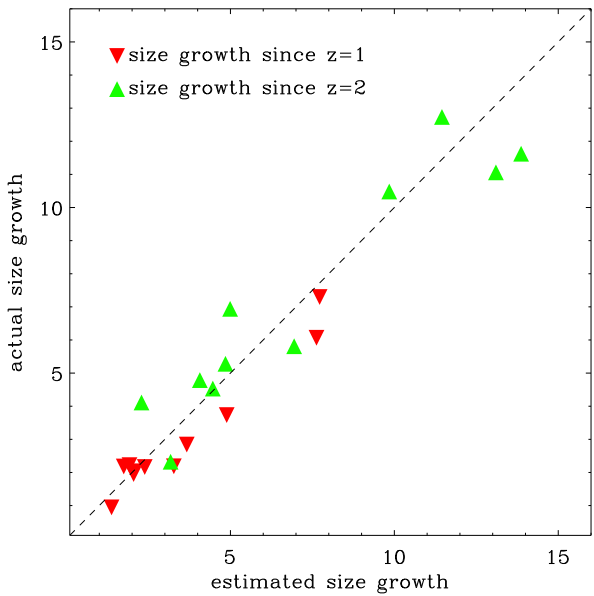}
\caption{The size growth predicted by equation \ref{rg} in combination
  with the stellar merger histories compared to the actual size growth
  in the simulations of the galaxies more massive than $M_* = 6.3
  \times 10^{10} M_{\odot}$ at z=2. The green triangles indicate the
  evolution between z=2 and z=0 the red triangles the evolution
  between z=1 and z=0. The simple virial estimate is a good predictor
  for the actual size evolution.}
\label{cmp-size-inc}
\end{figure}

\begin{figure*}
\centering
\includegraphics[width=18cm]{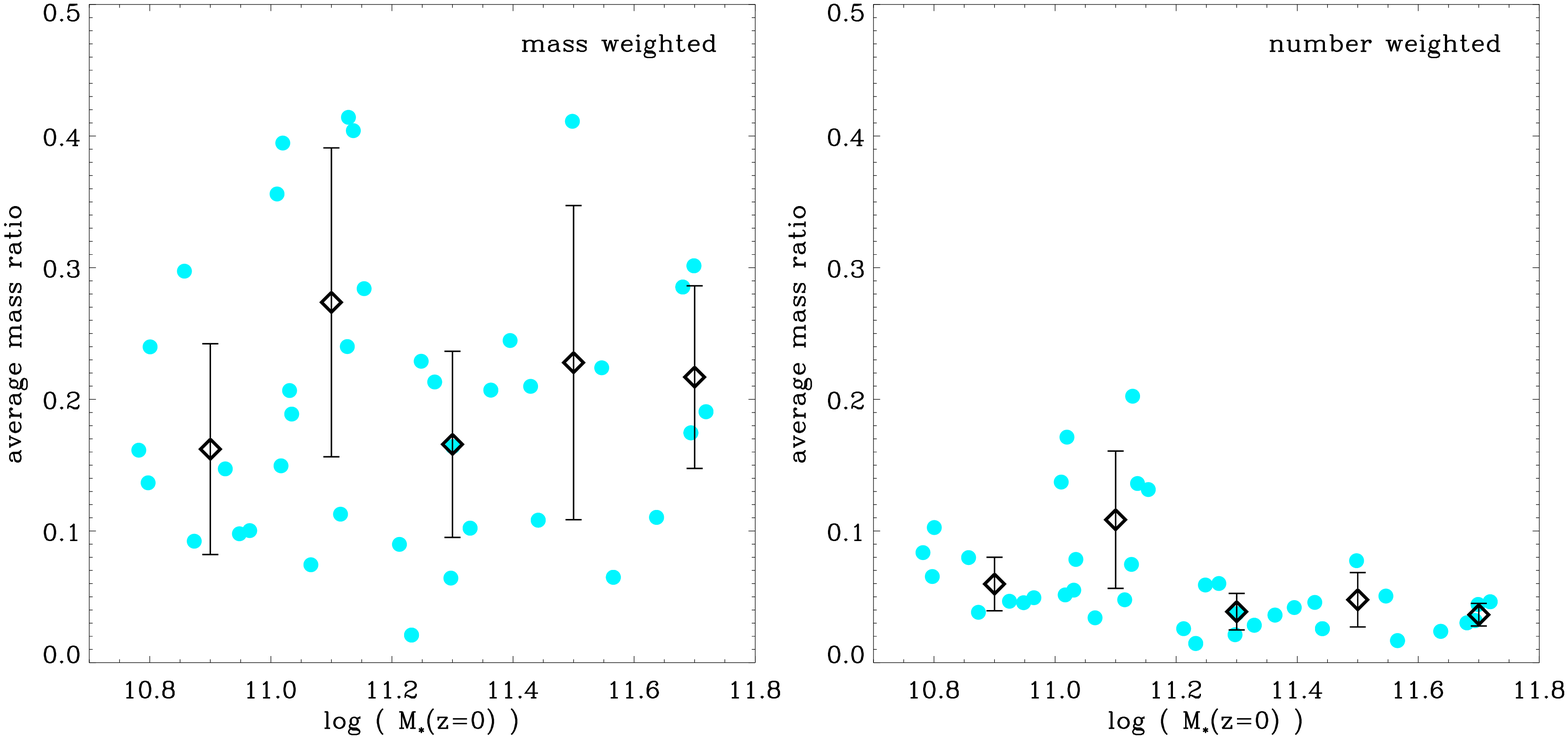}
\caption{Left: The average mass-weighted stellar merger mass-ratios
  (since z=2) as a function of present-day galaxy mass (blue
  dots). The black diamonds show the binned averages within 0.2 dex in
  stellar mass with the one sigma error bars. Trends with galaxy mass
  are statistically not significant. The mass growth is dominated by
  minor mergers with a mass ratio of $\approx$ 1:5. Right: The average
  number-weighted merger mass-ratio (for all stellar mergers since
  z=2) as a function of present-day galaxy mass. There is a weak trend
  for more massive galaxies to experience relatively more minor
  mergers. On average most stellar mergers have mass-ratio of $\approx
  1:16$.}
\label{mwmr}
\end{figure*}


\section{Stellar Merger histories}
\label{merger}

Similar to \citet{2000MNRAS.319..168C}, \citet{2009ApJ...699L.178N}
and \citet{2009ApJ...697.1290B} demonstrated, using the virial
theorem, how the size growth of a galaxy after a dissipationless
merger event can be predicted. Assuming energy conservation and
parabolic orbits \citep{2005MNRAS.358..551B, 2006A&A...445..403K} the
final gravitational radius of the system is given by
\begin{equation} 
\frac{R_{\mathrm{g,f}}}{R_{\mathrm{g,i}}} =
\frac{(1+\eta)^2}{(1+\eta\epsilon)}, \label{rg}
\end{equation}
where $R_{\mathrm{g,i}}$ and $R_{\mathrm{g,f}}$ are the initial and
final gravitational radii (before and after the merger event) which we
assume to be proportional to the spherical half mass radii
\citep{2009ApJ...699L.178N}.  Here, $\eta = M_\mathrm{a}/M_\mathrm{i}$
is the fractional mass increase during the merger and $\epsilon =
\langle \mathrm{v_a}^2 \rangle/\langle \mathrm{v_i}^2\rangle $ is the
ratio of the mean square speeds of the accreted satellites and the
initial system.  This estimate implies that the accretion of weakly
bound stellar systems ($\epsilon \ll 1$) results in a particularly
efficient size increase. To test the validity of this simple estimate
in our full cosmological simulation we follow the stellar merger
histories of our simulated galaxies. We identify every satellite
merging with the most massive progenitor of the central galaxy using a
FOF-finder with a minimum number of 20 stellar particles ($\approx 1.2
\times 10^{8} M_{\odot}$).  At $z=2$ all galaxies in our sample are
more massive than $\approx 10^{10.1}M_{\odot}$, i.e. we resolve
mergers at least down to a mass-ratio of $\approx 1:100$. For the
satellites as well as the host we compute the mass and the total
internal velocity dispersion and with this information and
Eqn. \ref{rg} we estimate the size evolution since z=2. In
Fig. \ref{cmp-size-inc} we show the estimated size growth for all
galaxies exceeding the z=2 mass limit of $6.3 \times 10^{10}M_{\odot}$
since redshift 2 (green triangles) and 1 (red triangles),
respectively. We compare this estimated size growth to the actual size
growth that we directly measure from our simulated
galaxies. Considering the simplifications used in Eqn. \ref{rg} -
homologous, one-component systems merging on zero energy orbits - the
predicted and actual growth agree notably well. This confirms earlier
findings \citep{2009ApJ...690.1452N} that the simple formula is a good
predictor even in a full cosmological context.  This approximation
however can only be valid if the assembly history for $z < 2$ is not
dominated by dissipational processes which is the case for the massive
systems presented here (see \citealp{2010ApJ...725.2312O} for the
relevant analysis).

To better understand the dominant assembly mechanism of our simulated
massive galaxies we have computed the average mass-weighted merger
mass-ratio for every galaxy since z=2. In the left panel of
Fig. \ref{mwmr} we show the average mass-weighted merger mass-ratio as
function of present-day stellar galaxy mass (blue dots). The
dependence on galaxy mass is weak. The average values in bins of 0.2
dex in mass are shown by the black diamonds with one sigma errors
bars. Overall the average mass-weighted merger mass-ratio is $\sim
0.20 \pm 0.10$. This makes 'minor mergers' with mass ratios of 1:5 the
dominant assembly mode, on average, for the massive simulated galaxies
{(see \citealp{2011arXiv1104.1626H} for a representation of a
  typical merger tree)}. The tendency of this ratio to change with the
mass of the host system cannot be determined by our calculations with
any statistical certainty (the slope of the fitted curve is $0.05 \pm
0.18$). However, we anticipate that for very low mass galaxies major
mergers would become more important. If the slope of the mass function
for satellites were $d(\ln \rm N)/ d(\ln \rm M) \sim -\gamma$, then
the expectation would be that the mass-weighted merger ratio would be
$(2-\gamma)/(3-\gamma) \sim 0.44$ if dynamical friction were not a
dominant process and $(3-\gamma)/(4-\gamma) \sim 0.64$ if it were
dominant. Thus, for low mass parent galaxies, we would anticipate that
the typical merger would be relatively 'major' with the ratio of
parent to satellite being $\sim 1:2$. Here we note that a significant
number of the simulated galaxies (7 out of 40, $\approx$ 18 per cent)
do not experience any merger with a mass ratio larger than 1:4,
e.g. they have experienced no major merger since z=2 at all.
{Estimates of merger rates for massive galaxies due to
  observations of disturbed systems
  \citep{2009ApJ...697.1971J,2011MNRAS.411.2148K,
    2011arXiv1108.2508L}, as well as } semi-analytic models lead to
similar results. E.g. \citet{2009MNRAS.397..506K},
\citet{2010ApJ...715..202H} and \citet{2010MNRAS.405..948S} find,
that massive early-type galaxies on average encounter less than one
major dry merger since their formation epoch. This confirms previous
suggestions motivated by the dearth of compact galaxies in the nearby
Universe, that a highly stochastical process like major mergers cannot
be the main driver for the observed size evolution
\citep{2009ApJ...697.1290B,2009ApJ...692L.118T,2010ApJ...720..723T}.
However, major mergers do happen and will have an impact on the
early-type galaxy population. {They can contribute
  significantly to the final stellar mass with minor mergers still
  dominating the size growth \citep{2010MNRAS.403..117S,
    2011arXiv1105.6043S}}. The observed merger rates, which are
difficult to determine, are in the range of only $\sim$ 1 major merger
since $z=2$
\citep{2006ApJ...640..241B,2009MNRAS.394L..51B,2006ApJ...636L..81N}
which is consistent with our interpretation.

In the right panel of Fig. \ref{mwmr} we show, the more conventionally
defined, average number-weighted merger mass-ratio. The merger history
since z=2 is clearly dominated by minor mergers with mass-ratios
smaller than 1:10. Those mergers, however, do on average not add most
of the mass to the systems. There is a slight trend for more massive
galaxies to experience a larger relative number of minor mergers. Over
the full mass range the average number-weighted merger mass ratio is
$\sim 0.062 \pm 0.043$, indicating that the typical merger was indeed
very minor (1:16).

\section{Conclusion \& Discussion} 
\label{discussion}

In this paper we use a sample of 40 cosmological re-simulations of
individual massive galaxies to investigate the evolution of galaxy
sizes and velocity dispersions with redshift. The simulated galaxies
form in a two phase process \citep{2010ApJ...725.2312O} where the
first phase at redshifts of $z \gtrsim 2$ is dominated by a
dissipative assembly. This formation phase is driven by in situ star
formation resulting in compact galaxies having small sizes of r
$\lesssim$ 1.3 kpc. The subsequent evolution of the galaxies at
redshifts of $z \lesssim 2$ is dominated by accretion of stars in
satellite stellar systems. \citet{2009ApJ...690.1452N} and
\citet{2010ApJ...725.2312O} have shown that the accreted stellar
systems preferentially settle into the outer parts of the galaxies,
resulting in a gradual increase in their sizes until the simulated
galaxies closely follow the present-day mass-size relation.  Between
redshift 2 and 0 our simulated galaxies grow on average by a factor of
$\sim 5-6$, whereas recent semi-analytical models find a smaller size
increase of $\sim 2-4$
\citep{2006ApJ...648L..21K,2011MNRAS.413..101G,2011arXiv1101.4225C}.  At the present day
25 out of the 40 simulated galaxies are quiescent ($\mathrm{sSFR} \leq
0.3 / t_{\mathrm{H}}$) and have structural parameters in agreement
with observed local early-type galaxies.  The underlying physical
reason for the size growth for our simulated galaxies is stellar
accretion, as can be seen in the strong positive correlation between
the projected stellar half-mass radii and the fraction of accreted
stellar material (Fig \ref{ia-rad-z}). Our detailed analysis presented
in this paper confirms that the stellar material is predominantly
accreted through minor mergers \citep{2009ApJ...699L.178N}, with
typical galaxy mass-ratios of $\approx 1:5$.  By number the merger
history is dominated by even more minor mergers with mass-ratios of
$\approx 1:16$. A significant fraction (18 per cent) of the galaxies
experience no major merger with mass-ratios larger than 1:4 since z=2
confirming previous suggestions, motivated by the lack of compact
galaxies in the nearby Universe, that a highly stochastical process
such as major mergers cannot be the main driver for the observed size
evolution
\citep{2009ApJ...697.1290B,2009ApJ...692L.118T,2010ApJ...720..723T}.
{Semi-analytical models also find significant stellar mass growth due
to minor mergers.  These models, however, predict that for the most
massive galaxies major mergers are becoming increasingly important
\citep[e.g.][]{1996MNRAS.283.1361B,
  2006MNRAS.366..499D,2007MNRAS.375....2D,2008MNRAS.384....2G}.  This
is a result of the sharp drop-off in the galaxy mass function due to
AGN feedback, which is not followed in our simulations.}

For galaxies with masses above $6.3 \times 10^{10}M_{\odot}$ our
simulated size evolution is in very good agreement (Fig
\ref{rad-gm-z0-shaded}) with the observed size evolution of galaxies
with similar masses at redshifts of $z \lesssim 2$
(e.g. \citealp{2008ApJ...688..770F}). The evolution of the sizes can
be well described by $R_{1/2} \propto (1+z)^{\alpha}$ with $\alpha =
-1.12$ for all galaxies and $\alpha = -1.44$ for quiescent galaxies
only.  The size growth measured from the simulations is in good
agreement with simple estimates from the virial theorem assuming
energy conservation during dissipationless merger events
\citep{2009ApJ...699L.178N}.

The projected velocity dispersions for simulated galaxies with masses
around $\approx 10^{11}M_{\odot}$ decrease systematically towards
lower redshifts from $\approx 262\; \rm{kms^{-1}}$ at z=2 to $177 \;
\rm{kms^{-1}}$ at $z=0$, again in good agreement with observations
(e.g. \citealp{2009ApJ...696L..43C}). Assuming an evolution as
$\sigma_{1/2} \propto (1+z)^{\beta}$ we find $\beta = 0.37$ for all
galaxies and $\beta = 0.44$ for quiescent galaxies. Future
observations might confirm this prediction.

We conclude that in the absence of dissipation and associated star
formation a growth scenario dominated by minor stellar mergers, with
less bound stars, is a viable physical process for explaining both the
observed growth in size and the decrease in velocity dispersion of
massive early-type galaxies from $z\sim 2$ to the present-day.
Accretion of systems not gravitationally bound to the central galaxy
causes, as noted, substantial size growth. But it has another,
dramatic, concomitant effect. As this mass becomes gravitationally
bound, it releases a large amount of gravitational energy.  This
process, which has been measured in our simulations
\citep{2008ApJ...680...54K,2009ApJ...697L..38J} , and termed
'gravitational heating' can add $\sim 10^{59.5} \rm ergs$ (i.e. $\sim
10^{43} \rm erg/s$) to the parent systems, causing heating of the
ambient gas and reducing the central dark matter component.

Despite these successes some obvious caveats concerning our
simulations remain. Most importantly our simulated galaxies are overly
efficient in transforming gas into stars \citep{2011arXiv1104.1626H}
and consequently the conversion efficiency of baryons into stars, even
at z=2, in the massive galaxies in our simulated sample is
overestimated by roughly a factor of $\approx$ 2 compared to
predictions from halo occupation models \citep[and references
  therein]{2010ApJ...717..379B}.  This discrepancy is most probably
due to the fact that our simulations neither include strong
supernova-driven winds nor AGN feedback from supermassive black holes
{and would be enhanced if metal-line cooling was
  included}. Observations and modeling have shown that strong galactic
winds generating significant outflows are ubiquitous at high redshifts
of $z\sim 2-3$ \citep[e.g.][]{2010ApJ...717..289S,
  2010arXiv1011.0433G} and in our simulations this aspect is missing
by construction. The effect of supernova driven winds and AGN feedback
is differential with respect to the masses of galaxies, with the
former primarily affecting smaller galaxies
\citep{2010MNRAS.406.2325O} and the latter being increasingly
important for more massive galaxies \citep{2009ApJS..182..216K}.
The {proper} inclusion of all the above mentioned physical
effects would certainly lower the overall total stellar masses
{(both the in-situ and the accreted component). Still,} the
relatively simple two-phase formation scenario provides a viable model
to physically explain the observed growth in size and decrease in
velocity dispersion. However, this is an issue clearly deserving
further studies on the effect of AGN feedback
\citep[e.g][]{2004MNRAS.347..144S, 2005Natur.433..604D,
  2005ApJ...620L..79S, 2009ApJ...690..802J, 2009MNRAS.398...53B,
  2010MNRAS.406L..55D, 2010ApJ...722..642O, 2010MNRAS.402.1536S},
radiative feedback from stars \citep[e.g][]{2006MNRAS.373.1265O,
  2009MNRAS.396.1383P, 2011arXiv1101.4940H} and feedback from
supernovae type II \citep[e.g][]{2008MNRAS.389.1137S,
  2010MNRAS.402.1536S, 2010MNRAS.409.1541S} and Ia
\citep[e.g][]{2007ApJ...665.1038C,
  2008MNRAS.387..577O}. Preferentially this will be investigated with
the help of a large sample of zoom simulations {with better
  statistics}, as presented here, with the aim of studying how these
processes would affect in detail the resulting size growth and
velocity dispersion evolution of massive galaxies.

\begin{acknowledgements}
We thank Ignacio Trujillo, Pieter van Dokkum, Simon D. M. White, and
Shardha Jogee as well as the anonymous referee for very useful
comments on the manuscript.  Part of the simulations were performed at
the Princeton PICSciE HPC center. This research was supported by the
DFG cluster of excellence 'Origin and Structure of the Universe' as
well as the DFG priority program 1177.

\end{acknowledgements}

\bibliography{./references.bib}

\end{document}